# EasyVitessce: auto-magically adding interactivity to Scverse single-cell and spatial biology plots

Selena Luo, Mark S. Keller, Tabassum Kakar, Lisa Choy, Nils Gehlenborg

Department of Biomedical Informatics, Harvard Medical School, Boston, MA 02115, United States

## Abstract

**Summary:** EasyVitessce is a Python package that turns existing static Scanpy and SpatialData plots into interactive visualizations by virtue of adding a single line of Python code. The package uses Vitessce internally to render interactive plots, and abstracts away technical details involved with configuration of Vitessce. The resulting interactive plots can be viewed in computational notebook environments or their configurations can be exported for usage in other contexts such as web applications, enhancing the utility of popular Scverse Python plotting APIs.

**Availability and implementation:** EasyVitessce is released under the MIT License and available on the Python Package Index (PyPI). The source code is publicly available on GitHub (https://github.com/vitessce/easy_vitessce). Documentation with examples is provided at https://vitessce.github.io/easy_vitessce/.

**Contact:** Nils Gehlenborg, nils@hms.harvard.edu

## Introduction

The single-cell data analysis ecosystem is large (Zappia et al. 2018), yet the community has converged on a small set of core data structures and analysis packages. These enable interoperability, streamline common tasks, and facilitate software maintenance. In Python, the Scverse project maintains the AnnData (Virshup et al. 2024), MuData (Bredikhin et al. 2022), and SpatialData data structures and several analysis packages, including the most widely-used, Scanpy (Wolf *et al.*, 2018). Visualization and the creation of static figures comprises a key part of the Scanpy API, with functions to generate scatterplots, heatmaps, dot plots, among other chart types. Scanpy is typically imported into Python scripts such that the "sc.pl" (plotting) shortcut is available to the analyst with only a few key presses. For spatially-resolved data that is stored in SpatialData objects, the SpatialData-Plot package similarly enables a ".pl" shortcut to plot each type of spatial element (shapes, points, images, and labels) (Marconato *et al.*, 2025).

While the usage of specialized data structures and analysis packages enables complex static plotting operations to be performed with very few lines of code, there is currently a barrier to generation of interactive plots using the same syntax. In other words, transitioning from a static plot to an interactive plot currently requires using a separate package with its own set of plotting APIs (e.g., function and parameter names) that may require extensive code modification.

There are numerous benefits of interactive plotting in the context of single-cell data analysis. For instance, lasso interactions in scatterplots enable selection of cells of interest, which can be used for downstream analysis such as subclustering. In plots of spatial and imaging data, interactions such as zooming and panning enable smooth navigation from overview-to-detail. In all plots that display one biomarker (or a limited number of them) at a time, interactivity can enable users to easily select alternative biomarker(s) to view their expression patterns.

To achieve the best of both worlds (static and interactive), we developed EasyVitessce, a Python package for the creation of interactive single-cell and spatial visualizations via existing static plotting APIs available from the Scanpy and SpatialData-Plot packages. EasyVitessce enables seamless transition between static and interactive plotting, as it requires only adding a single line of code: to import the package, which enables interactive variants of all plots by default, with options to disable.

## Methods

EasyVitessce uses the interactive plotting capabilities of Vitessce [(Keller et al. 2024)](#), a scalable framework for spatial and multi-modal single-cell data visualization. Specifically, EasyVitessce relies on the Vitessce Python package, adopting its features such as support for multiple computational environments including Jupyter, JupyterLab, Marimo, and Google Colab notebooks [(Manz et al., 2024)](#). The Vitessce Python package also contains utilities for serving local files in order to visualize them in the Vitessce notebook widget.

To guide the development of EasyVitessce, we first conducted a code search using the GitHub API to survey plotting function usage for Scanpy and SpatialData, both part of the Scverse ecosystem [(Virshup et al. 2024; Marconato et al. 2025)](#). This provided us with an understanding of which plotting functions to prioritize, based on their usage in open-source code (Supplemental Information). Plotting functions not currently supported by Vitessce, such as the dendrogram plot, are not included in EasyVitessce and thus the behavior of these functions is unchanged by EasyVitessce.

In order to streamline the process of transitioning between static and interactive plotting, the EasyVitessce function signatures match those of the plotting functions from Scanpy and SpatialData-Plot. We used a software engineering technique called "monkeypatching" to swap out the plotting function definitions from these packages at runtime. This approach enables users to retain their existing plotting function calls, syntax, and mental models. Other methods we considered were swapping out the "backend" for plotting, similar to Matplotlib's implementation[1]. This would require close coordination with Scanpy's developers, and remains a future direction for the project. Another option would be to require the user to modify their plotting code (e.g., to import plotting functions from the EasyVitessce package explicitly). Although this would circumvent the need for monkeypatching, the user would need to modify every plotting function call, preventing easy transition between static and interactive plotting (and vice-versa). Thus, to maintain our goal of requiring only one additional line of code, our implementation matches the syntax of existing plotting functions through monkeypatching. While EasyVitessce improves interoperability between Vitessce and ScanPy/SpatialData-Plot, the broader question of how to enable interoperability across all visualization packages in the Scverse ecosystem remains.

## Results

The EasyVitessce approach enables users to create interactive figures while retaining the existing, widely-adopted Scverse plotting APIs. Not only can interactive figures be created, but existing static figures can be seamlessly transitioned to interactive figures through the addition of a single line of code. EasyVitessce supports nine Scanpy plotting functions (e.g., sc.pl.dotplot) and all four plotting functions of SpatialData-Plot (e.g., sdata.pl.render_shapes). The list of plotting functions and parameters that are currently supported by EasyVitessce can be found in the Supplementary Information.

## Conclusion

EasyVitessce is a Python package for streamlining the process of transitioning between static and interactive plots. The package supports frequently used plotting functions from Scanpy and SpatialData, and preserves their syntax through a monkeypatching technique. Rather than necessitating a new block of code in order to add interactivity to a static plot, this approach allows the user to keep their existing line of plotting code unchanged. The package supports parameters for these plotting functions, which allow the user to customize the visual encodings and properties. Future work includes development of an EasyVitessce package for R using an analogous approach based on existing plotting functions from the Seurat ecosystem (Satija et. al, 2015), as well as functionalities to combine interactive plots returned by EasyVitessce into a larger grid of coordinated subplots to support creation of complex multi-view visualizations.

---

[1] https://matplotlib.org/stable/users/explain/figure/backends.html

# Figures

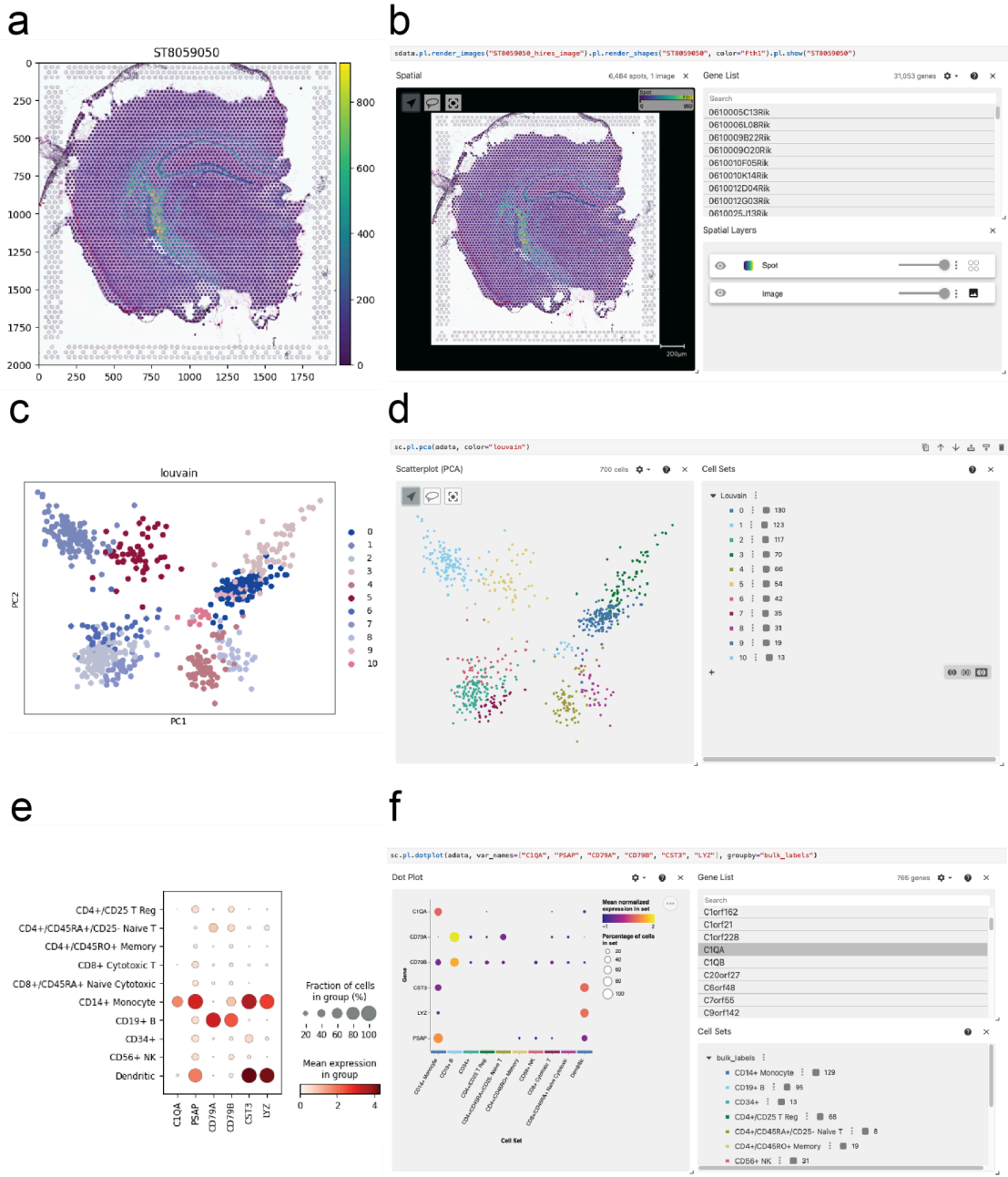

*Fig. 1: Plots before and after importing EasyVitessce. A) Usage of the SpatialData-Plot API to visualize a 10x Genomics Visium mouse brain dataset, rendering both its histology image and overlaid spots colored by the expression of the gene Fth1. B) The same plotting code used in (A), with EasyVitessce imported to enable interactive plotting, results in a Vitessce widget output. C) Usage of the Scanpy plotting API to view PCA coordinates and Louvain clustering results for a PBMC example dataset (loaded via the Scanpy datasets API). D) The same plotting code used in (C), with EasyVitessce enabled, results in a Vitessce widget output. E) Usage of the Scanpy plotting API to generate a dot plot. F) With EasyVitessce, the same code to generate the static*

*dot plot from (E) results in an interactive output. (Note that Vitessce uses slightly different default styling and colormaps (e.g., the dot plot in Vitessce is transposed and uses a plasma colormap). These styling differences could be minimized through future updates to the Vitessce core library.)*

## Funding

This project was supported by the HuBMAP Summer Internship Program, and NIH awards U24 CA268108, OT2 OD033758, and R33 CA263666.